\documentclass[aps,pre,showpacs,groupedaddress]{revtex4}
\usepackage{amsfonts}
\usepackage{mathrsfs}
\usepackage{graphicx}
\usepackage{amsmath}
\usepackage{amssymb}
\usepackage{amsbsy}

\begin{document}

\title{Reconstructing the Hopfield network as an inverse Ising problem}

\author{Haiping Huang}
\affiliation{Key Laboratory of Frontiers in Theoretical Physics,
Institute of Theoretical Physics, Chinese Academy of Sciences,
Beijing 100190, China}
\date{\today}

\begin{abstract}
We test four fast mean field type algorithms on Hopfield networks as
an inverse Ising problem. The equilibrium behavior of Hopfield
networks is simulated through Glauber dynamics. In the low
temperature regime, the simulated annealing technique is adopted.
Although performances of these network reconstruction algorithms on
the simulated network of spiking neurons are extensively studied
recently, the analysis of Hopfield networks is lacking so far. For
the Hopfield network, we found that, in the retrieval phase favored
when the network wants to memory one of stored patterns, all the
reconstruction algorithms fail to extract interactions within a
desired accuracy, and the same failure occurs in the spin glass
phase where spurious minima show up, while in the paramagnetic
phase, albeit unfavored during the retrieval dynamics, the
algorithms work well to reconstruct the network itself. This implies
that, as a inverse problem, the paramagnetic phase is conversely
useful for reconstructing the network while the retrieval phase
loses all the information about interactions in the network except
for the case where only one pattern is stored. The performances of
algorithms are studied with respect to the system size, memory load
and temperature, sample-to-sample fluctuations are also considered.

\end{abstract}

 \maketitle

\section{Introduction}
Last years have witnessed a surge of research interests in the
inverse Ising problem \cite{Bialek-06ep, Bialek-07ep, Mezard-08,
SM-09, Roudi-2008, Roudi-2009, Roudi-2009ep, Marre-09, Cocco-09},
also known as Boltzmann machine learning in statistical inference
theory \cite{Ack-1985, Hertz-1991}. This is due to on one hand the
fact that a huge amount of data can be collected from many
biological systems such as neural networks, gene regulatory networks
and metabolic networks \cite{Shlens-2006, Nature-06, Tang-2008,
Lezon-06, Vert-08}, on the other hand to the growing need for novel
and efficient algorithms to reconstruct the network based on the
huge amount of experimental data. Individual elements (e.g.,
neurons, genes, even computers in the internet) in the network
usually interact with each other to yield the collective behavior
emerging at the network level. To extract functional connectivity
from the collective behavior of the network, we use
$\{\sigma_{i}\}_{i=1}^{N}$ to represent the activity (e.g., electric
activity of single neuron, expression level of single gene) of each
element in a network with size $N$, then the likelihood of each
state $\boldsymbol{\sigma}$ is assumed to be $P_{{\rm
Ising}}(\boldsymbol{\sigma})\propto\exp\left[\sum_{i<j}J_{ij}\sigma_{i}\sigma_{j}+\sum_{i}h_{i}\sigma_{i}\right]$,
also named the second-order maximum entropy model studied in
Refs.~\cite{Nature-06, Tang-2008, Lezon-06}. $\{h_{i},J_{ij}\}$
serve as Lagrange multipliers corresponding to the constraints given
by $\{m_{i},C_{ij}\}$ where $m_{i}$ is the magnetization and
$C_{ij}$ two-point connected correlation between sites $i$ and $j$
in the statistical physics language. From the experimental data
(e.g., microarray data or multi electrode recordings), one can
measure both the mean activity ($m_{i}$) of each individual and
pairwise correlations ($C_{ij}$) among them. The inverse Ising
problem is to infer the underlying parameters $\{h_{i},J_{ij}\}$
from the knowledge of measured $\{m_{i},C_{ij}\}$, such that the
resulting Ising distribution is able to provide an accurate
description of the statistics of the experimental data, i.e.,
$\left<\sigma_{i}\right>_{{\rm Ising}}=\left<\sigma_{i}\right>_{{\rm
data}}, \left<\sigma_{i}\sigma_{j}\right>_{{\rm
Ising}}=\left<\sigma_{i}\sigma_{j}\right>_{{\rm data}}$.

The pairwise Ising model has been extensively studied as an inverse
Ising problem on retinal networks~\cite{Shlens-2006,Nature-06},
cortical networks~\cite{Tang-2008} and gene regulatory
networks~\cite{Lezon-06}. In Ref.~\cite{Lezon-06}, it was observed
that the maximum entropy principle can be used to extract
information about gene interactions and the result reproduces the
observed transcript profiles with high fidelity. Schneidman
\textit{et al} also showed that the pairwise Ising model captures
$\sim90\%$ of the correlation structure of the retinal network
activity. On top of the existing spatial correlations among neurons,
the temporal dependencies were also suggested to be a common feature
of cortical networks~\cite{Tang-2008}. Recently, Marre \textit{et
al} employed the same maximum entropy principle with a Markovian
assumption to predict the occurrence probabilities of spatiotemporal
patterns and the result is significantly better than that obtained
by Ising models only considering spatial correlations. In aspects of
theoretical analysis, Roudi \textit{et
al}~\cite{Roudi-2008,Roudi-2009ep} have investigated the dependence
of the fit quality of pairwise model upon the time bin size as well
as the system size and found that the pairwise model always provides
an accurate statistical description of spikes as long as the system
size does not exceed the critical size determined by the mean
population firing rate and the bin size. From the algorithmic
perspective, Broderick \textit{et al}~\cite{Bialek-07ep} combined a
coordinate descent algorithm with an adaption of the histogram Monte
Carlo method to solve the inverse problem efficiently up to $N=40$
neurons. Subsequently, M\'ezard and Mora introduced the message
passing ideas to the inverse Ising problem and proposed the
susceptibility propagation as a comprehensive network reconstruction
algorithm for sparse network or the network with sufficiently weak
interactions~\cite{Mezard-08}. The susceptibility propagation
together with Sessak-Monasson approximation (SM) recently put
forward in Ref.~\cite{SM-09}, has been tested in
Sherrington-Kirkpatrick model~\cite{Mezard-1987}, and it was found
that the message passing based method as well as SM outperforms
other existing mean field schemes and SM was shown to be more
efficient. The message passing technique also found application in
the inference of gene regulatory networks, and a statistical
mechanics analysis has been presented in Ref. ~\cite{Braunstein-08}.
Roudi \textit{et al} in a recent work~\cite{Roudi-2009} studied the
inverse problem on a simulated network of spiking neurons and it was
observed that as the network size increases, SM and the inversion of
Thouless-Anderson-Palmer (TAP) equations outperform other mean field
type algorithms to predict the network structure yet the fit quality
degrades as the system size grows.

In this paper we use the same four mean field schemes employed in
Ref.~\cite{Roudi-2009} to test their performances on both the
fully-connected and finite connectivity Hopfield network and
investigate the behavior of these inference algorithms with respect
to the system size, the memory load and particularly different
temperatures. It is shown that the paramagnetic phase helps to
extract the information about couplings in the network, although the
recall phase is in turn useful during the retrieval process of one
of embedded patterns, and in this case the system is prevented from
entering the paramagnetic or spin glass phase. The naive mean field
method (nMF) and TAP exhibit very good performances while
independent-pair approximation (ind) shows a relatively high
inference error. In the paramagnetic phase, SM leads to the nearly
identical performances with nMF and TAP, and their performances
deteriorate with increasing network size. Given the fixed system
size, all the algorithms show increasing inference errors as the
memory load increases. In the low temperature region, SM shows
relatively high inference errors and is even inferior to ind for
certain memory loads. The sample-to-sample fluctuations are also
taken into account, and we found the inference error shows large
fluctuations as the temperature decreases. The finite connectivity
Hopfield network is also analyzed, as expected, ind performs well
especially for the small number of stored patterns or at low
temperature. In addition, when the temperature decreases, the
reconstruction algorithms show similar behaviors as observed in the
fully-connected network.

The remainder of this paper is organized as follows. The Hopfield
model and Glauber dynamics (GD) are introduced in
Sec.~\ref{sec_Hopf}. In Sec.~\ref{sec_MFS}, four mean field schemes,
nMF, ind, SM and TAP, are presented briefly. The reconstruction
performances of these algorithms are reported in
Sec.~\ref{sec_result} for the fully-connected network and the finite
connectivity network respectively. We conclude with our results and
future perspectives in Sec.~\ref{sec_Con}.

\section{Hopfield Networks and Glauber Dynamics}
\label{sec_Hopf}

The Hopfield network, proposed in Ref.~\cite{Hopfield-1982}, later
thoroughly discussed in Refs.~\cite{Amit-198501,Amit-198502},
functions as an associative memory network. It is in essence a
recurrent network, and its equilibrium properties are determined by
the following Hamiltonian:
\begin{equation}\label{Hop-H}
    \mathcal{H}=-\frac{1}{2}\sum_{i\neq j}J_{ij}\sigma_{i}\sigma_{j}
\end{equation}
where $\sigma_{i}$ represents the state of each neuron in the
network. $\sigma_{i}=+1$ indicates the neuron $i$ generates a spike
while $\sigma_{i}=-1$ keeps quiescent. Interactions between neurons
are constructed according to the Hebb's rule,
\begin{equation}\label{Hop-J}
    J_{ij}=\frac{1}{N}\sum_{\mu=1}^{P}\xi_{i}^{\mu}\xi_{j}^{\mu}
\end{equation}
where $\{\xi_{i}^{\mu}\}$ taking $\pm1$ with equal probability
$\frac{1}{2}$ are $P$ stored patterns. The number of patterns $P$
scales as $P=\alpha N$ in the fully-connected case. The Hebb's rule
expresses the multiplicative interaction between presynaptic and
postsynaptic activity and positively correlation (both neurons are
on or off) causes an enhanced coupling while negatively correlation
results in a decreased one. Under the Hebb's rule, the Hamiltonian
equation ~(\ref{Hop-H}) can be actually rewritten in terms of the
overlap between the network configuration and the stored patterns.
This guarantees that the energy function $\mathcal{H}$ always
decreases while the system evolves according to the following GD
rule.

The fully-connected Hopfield network requires complete and symmetric
connectivity, also no self-interactions. Mean field behavior of
finite connectivity Hopfield network has been recently studied in
Refs.~\cite{Coolen-2003,Skantzos-2004}. The difference is that in
the sparse network, the coupling $J_{ij}$ is constructed as follows,
\begin{equation}\label{Hop-J-sparse}
    J_{ij}=\frac{l_{ij}}{l}\sum_{\mu=1}^{P}\xi_{i}^{\mu}\xi_{j}^{\mu}
\end{equation}
where $l$ is the mean degree of each neuron. When
$N\rightarrow\infty$, $P=\alpha l$. We also assume that no
self-interactions are present in the finite connectivity Hopfield
model and the connectivity $l_{ij}$ is symmetric and subject to the
distribution
\begin{equation}\label{distri}
    P(l_{ij})=(1-\frac{l}{N-1})\delta(l_{ij})+\frac{l}{N-1}\delta(l_{ij}-1)
\end{equation}

Taking the thermal fluctuation into account, the GD rule
~\cite{Glauber-1963} is specified as
$P(\sigma_{i}\rightarrow-\sigma_{i})=\frac{1}{1+\exp(\beta\Delta\mathcal{H}_{i})}$
where $\Delta\mathcal{H}_{i}$ stays for the energy change due to
such a flip. Equivalently, the dynamics rule can be recast
into~\cite{Biroli-1998},
\begin{equation}\label{GDrule}
    P(\sigma_{i}\rightarrow-\sigma_{i})=\frac{1}{2}\left[1-\sigma_{i}\tanh\beta h_{i}\right]
\end{equation}
where the inverse temperature $\beta$ serves as a measure of degree
of stochasticity and $h_{i}=\sum_{j\neq i}J_{ij}\sigma_{j}$ is the
local field acting on $\sigma_{i}$. Under this dynamics rule, if the
initial configuration is close enough to one of embedded patterns,
it will tend to evolve towards the nearest attractor represented by
one single pattern in the configuration space. That is also the
meaning of associative memory.

In this work, we use GD to detect the equilibrium properties of
Hopfield networks, then the numerical simulation data is used to
reconstruct the network. We shall keep to the small size system up
to $N=190$ due to the computational cost. In the thermodynamic
limit, the phase diagram of fully-connected Hopfield network has
been studied in Refs.~\cite{Amit-198501,Amit-198502} while that of
sparse network in Refs.~\cite{Coolen-2003,Skantzos-2004}. In the
numerical simulation, two types GD will be implemented. Both types
are run in a randomly asynchronous manner. The type ${\rm A}$ is
executed as follows. We do the standard GD as demonstrated in
equation~(\ref{GDrule}) totally $4\times 10^{6}$ steps (each step
corresponds to the process where the state of each neuron is updated
on average one time), among which the first $2\times 10^{6}$ steps
are run for the system to reach the equilibrium state and the other
$2\times 10^{6}$ steps for calculating magnetizations and
correlations. We sample the state of the network every $200$ steps.
To get around the difficulty that the system is apt to get stuck in
local minima of the free energy landscape at low temperature, the
simulated annealing technique~\cite{Mezard-1987} is introduced in
type ${\rm B}$ GD where we set the initial temperature to be $1.0$
and the cooling rate $0.005$. At each intermediate temperature, we
run GD $10^{4}$ steps. When the temperature is decreased to the
desired one, we run another $2\times 10^{6}$ steps to sample the
system. The smaller cooling rate and larger steps run at each
intermediate temperature are favored but the corresponding
computational cost increases.

\section{Mean Field Schemes for Inferring Couplings}
\label{sec_MFS}

Boltzmann machine learning works also as a network reconstruction
algorithm~\cite{Ack-1985,Hertz-1991}. It adjusts couplings at each
iteration step according to the difference between measured
correlation and that produced by the prescribed Ising distribution.
The algorithm is iterated until the difference falls within the
desired accuracy. The Boltzmann machine learning is exact but also
computationally expensive since a large amount of Monte Carlo
sampling steps are required. In this section, we will briefly
introduce for the Hopfield network reconstruction four fast mean
field approximations presented in Ref.~\cite{Roudi-2009}.

\subsection{Naive Mean-Field Method}
\label{subsec:nMF}

The naive mean field theory indicates that
$m_{i}=\tanh\left(h_{i}+\sum_{k\neq i}J_{ik}m_{k}\right)$ where
$m_{i}=\left<\sigma_{i}\right>$. To calculate the connected
correlation $C_{ij}=\left<\sigma_{i}\sigma_{j}\right>-m_{i}m_{j}$,
we use the fluctuation-response relation,
\begin{equation}\label{corre}
    C_{ij}=\frac{\partial m_{i}}{\partial
    h_{j}}=(1-m_{i}^{2})\left[\delta_{ij}+\sum_{k\neq i}J_{ik}C_{kj}\right]
\end{equation}
then obtain the nMF prediction of couplings,
\begin{equation}\label{nmf}
    J_{ij}^{{\rm nMF}}=(\mathbf{P}^{-1})_{ij}-(\mathbf{C}^{-1})_{ij}
\end{equation}
where $\mathbf{P}_{ij}=(1-m_{i}^{2})\delta_{ij}$. Note that the
predicted $J_{ij}$ here has been multiplied by $\beta$, therefore
the actual predicted one $J_{ij}^{a}=J_{ij}/\beta$.

\subsection{Independent-Pair Approximation}
\label{subsec:Ind}

This approximation assumes that each pair of neurons are independent
of the rest part of the system, i.e., their joint probability
$P(\sigma_{i},\sigma_{j})\propto\exp\left[h_{i}^{(j)}\sigma_{i}+h_{j}^{(i)}\sigma_{j}+J_{ij}\sigma_{i}\sigma_{j}\right]$
where $h_{i}^{(j)}(h_{j}^{(i)})$ is the local field neuron $i(j)$
feels when neuron $j(i)$ is removed from the system. Then the ind
prediction is given by
\begin{equation}\label{ind}
    J_{ij}^{{\rm ind}}=\frac{1}{4}\log\left[\frac{(1+C_{ij}^{'})^{2}-(m_{i}+m_{j})^{2}}{(1-C_{ij}^{'})^{2}-(m_{i}-m_{j})^{2}}\right]
\end{equation}
where $C_{ij}^{'}=C_{ij}+m_{i}m_{j}$.

\subsection{Sessak-Monasson Approximation}
\label{subsec:SM}

In a recent work by Sessak and Monasson~\cite{SM-09}, the SM
prediction of couplings is derived using a systematic small
correlation expansion,
\begin{equation}\label{SM}
    J_{ij}^{{\rm SM}}=J_{ij}^{{\rm loop}}+J_{ij}^{{\rm ind}}-\frac{C_{ij}}{(1-m_{i}^{2})(1-m_{j}^{2})-C_{ij}^{2}}
\end{equation}
where $J_{ij}^{{\rm ind}}$ is given by equation ~(\ref{ind}) and
$J_{ij}^{{\rm
loop}}=(L_{i}L_{j})^{-1/2}\left[\mathbf{M}(\mathbf{I}+\mathbf{M})^{-1}\right]_{ij}$
where $\mathbf{I}$ is an identity matrix, $L_{i}=1-m_{i}^{2},
\mathbf{M}_{ij}=C_{ij}(L_{i}L_{j})^{-1/2}$ and $\mathbf{M}_{ii}=0$.

\subsection{Inversion of TAP Equations}
\label{subsec:InvTAP}

The usual TAP equation reads $h_{i}=\tanh^{-1}m_{i}-\sum_{j\neq
i}J_{ij}m_{j}+m_{i}\sum_{j\neq
i}J_{ij}^{2}(1-m_{j}^{2})$~\cite{Mezard-1987}. Taking the derivative
of the field $h_{i}$ with respect to the magnetization $m_{j}$, one
readily obtains the TAP prediction equation,
\begin{equation}\label{TAP}
    (\mathbf{C}^{-1})_{ij}=\frac{\partial h_{i}}{\partial m_{j}}=-J_{ij}^{{\rm
    TAP}}-2(J_{ij}^{{\rm TAP}})^{2}m_{i}m_{j}
\end{equation}
which has been introduced by Kappen and Rodriguez~\cite{Kappen-1998}
and Tanaka~\cite{Tanaka-1998}.

To measure the reconstruction performance of these algorithms, we
define the inference error as
\begin{equation}\label{inferror}
    I_{e}=\left[\frac{2}{N(N-1)}\sum_{i<j}(J_{ij}^{{\rm pred}}-J_{ij}^{{\rm true}})^{2}\right]^{1/2}
\end{equation}
where $J_{ij}^{{\rm pred}}$ is the prediction value of the coupling
based on the above four reconstruction algorithms and $J_{ij}^{{\rm
true}}$ the original coupling constructed through the Hebb's rule
equation ~(\ref{Hop-J}) or equation ~(\ref{Hop-J-sparse}).
Obviously, the smaller $I_{e}$ is, the more precisely the pairwise
Ising model reproduces the statistics of the experimental data.

\begin{widetext}
\begin{center}
\begin{figure}
    \includegraphics[width=8.5cm]{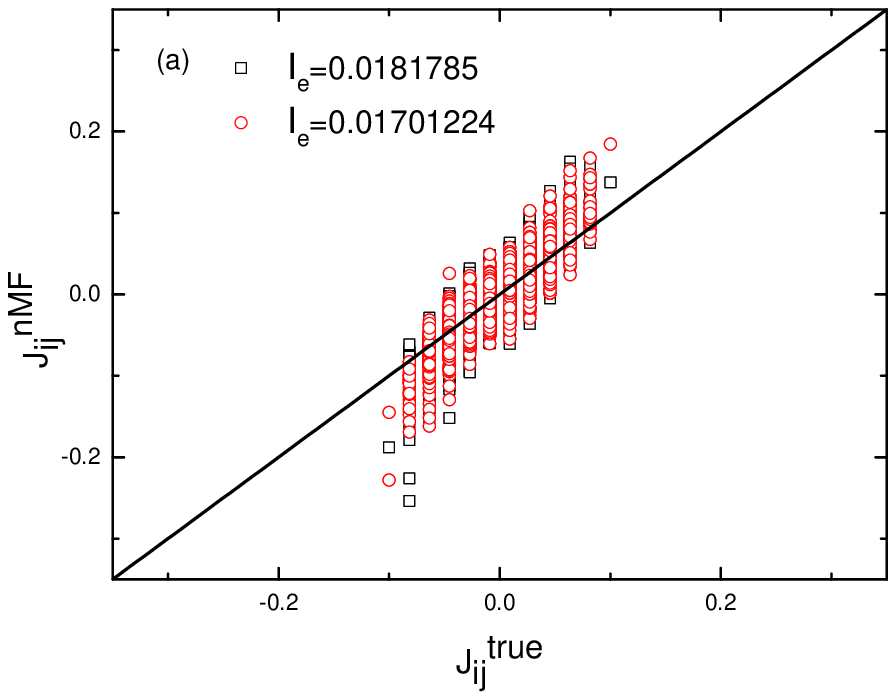}
    \hskip .5cm
    \includegraphics[width=8.5cm]{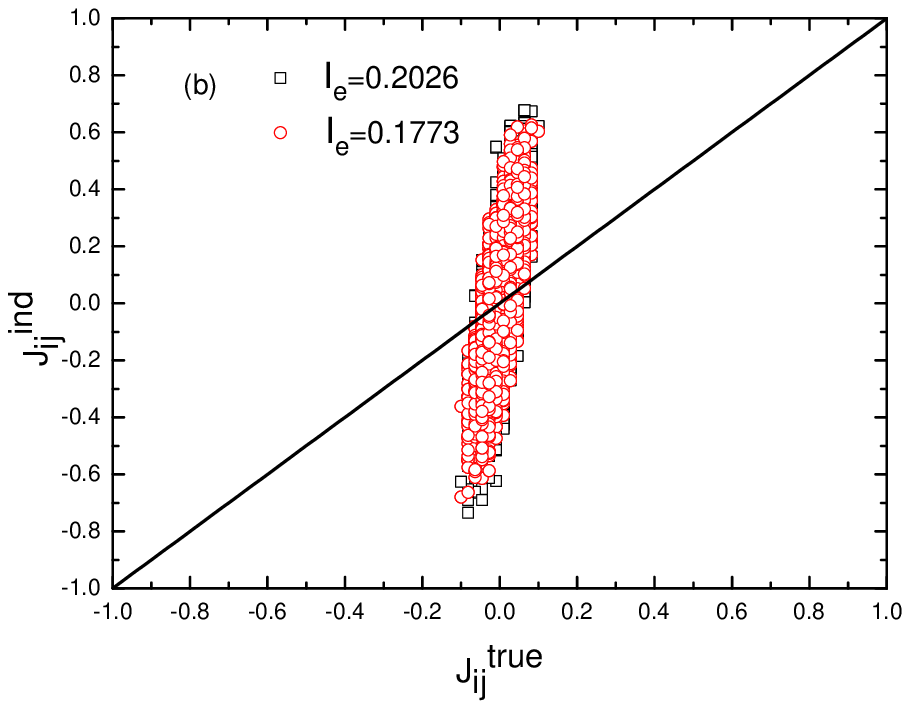}
    \vskip .2cm
    \includegraphics[width=8.5cm]{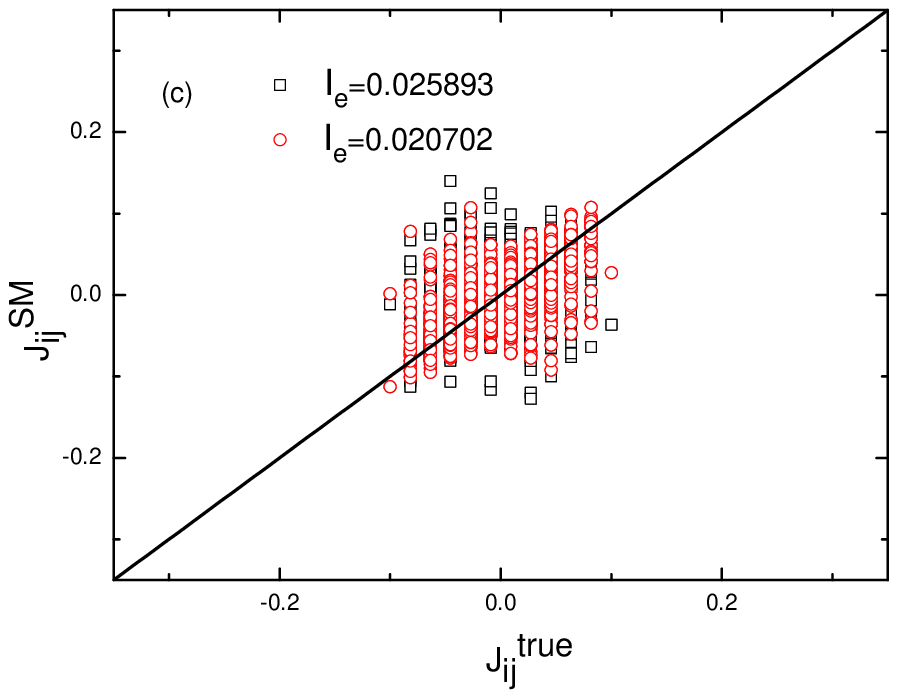}
     \hskip .5cm
    \includegraphics[width=8.5cm]{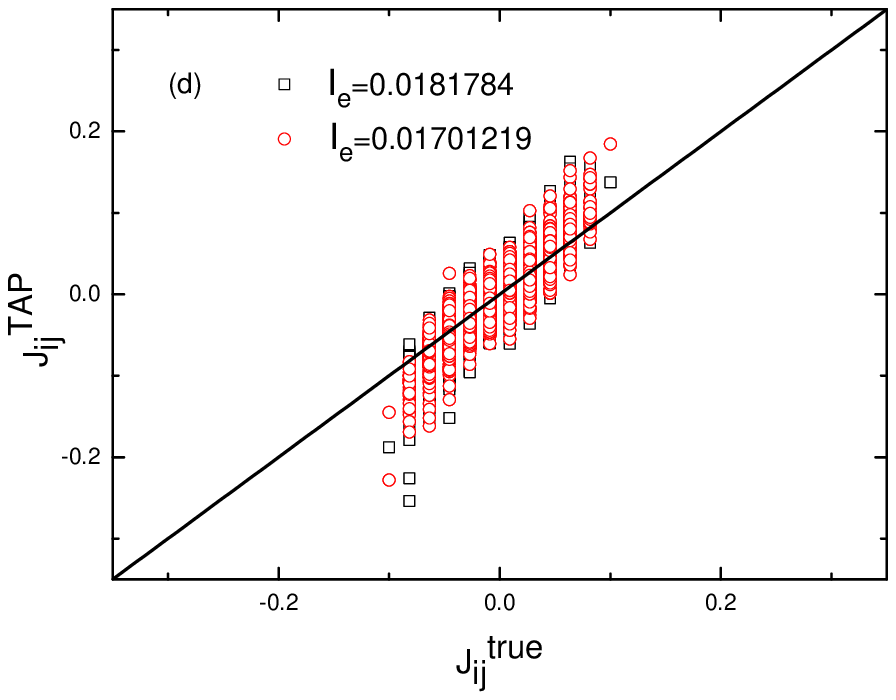}\vskip .2cm
  \caption{
    (Color online)  Scatter plots comparing the inferred couplings with the true
    ones for $\beta=1.0, \alpha=0.1, N=110$. Results from two samples are shown respectively. The full line indicates equality and the network is
    fully-connected. (a) nMF, (b) ind,
    (c) SM and (d) TAP.
  }\label{SG_scatterplot}
\end{figure}
\end{center}
\end{widetext}

\begin{widetext}
\begin{center}
\begin{figure}
    \includegraphics[width=8.5cm]{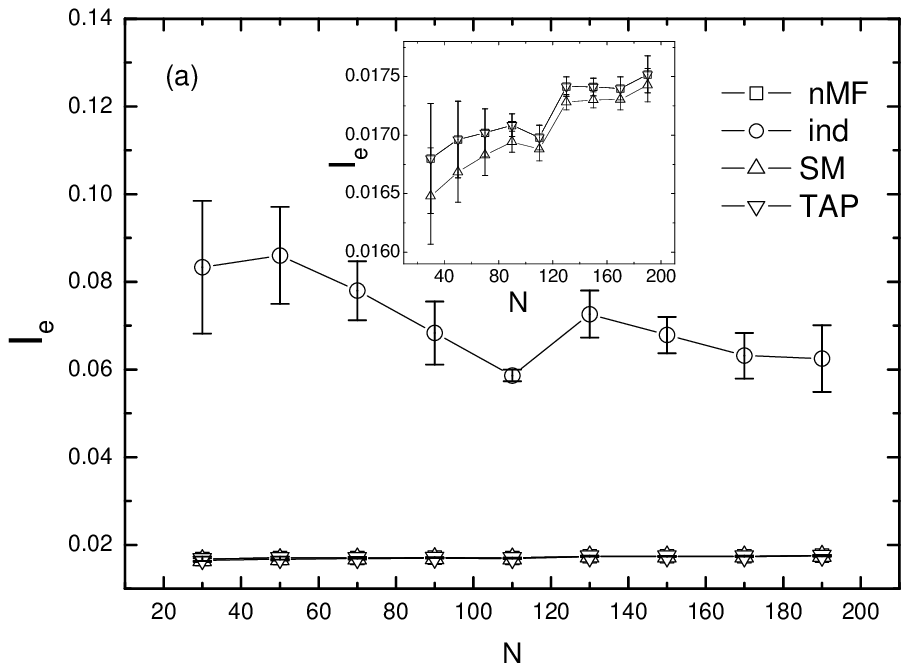}
    \hskip .5cm
    \includegraphics[width=8.5cm]{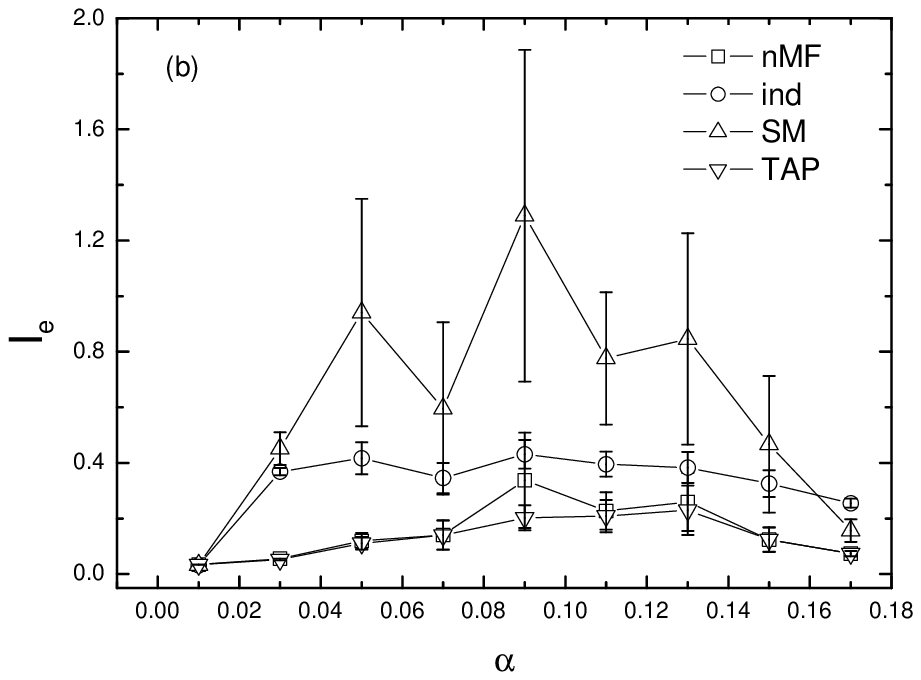}
    \vskip .2cm
    \includegraphics[width=8.5cm]{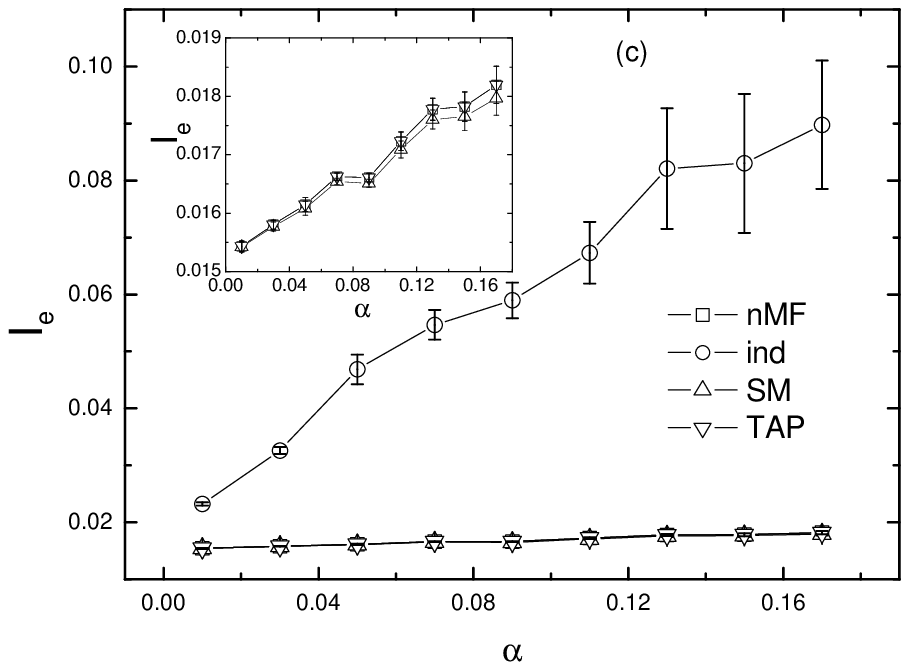}
    \hskip .5cm
     \includegraphics[width=8.5cm]{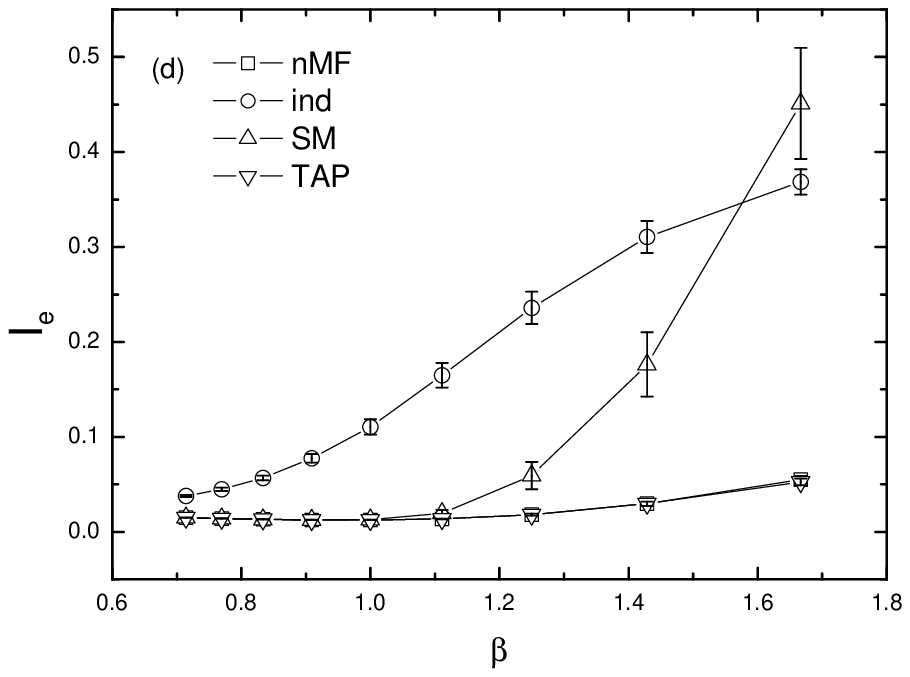}\vskip .2cm
  \caption{
     Inference performances for the fully-connected Hopfield network. The line linking
    each marker is a guide to the eye. The data marker is the average over $5$ samples. (a) The inference error against the system size for the paramagnetic phase
    $T=1.5, \alpha=0.1$. Results are obtained based on type ${\rm A}$ GD. The inset shows an enlarged view of performances
    for nMF, SM as well as TAP. (b) The inference error against the memory
load with $T=0.6, N=100$. Results
    are obtained based on type ${\rm B}$ GD. (c) The inference error against the memory
load with $T=1.5, N=100$. Results
    are obtained based on type ${\rm A}$ GD. The inset shows an enlarged view of performances
    for nMF, SM and TAP. (d) The inference error
    versus temperature for $\alpha=0.03, N=100$. For the low
    temperature regime, type B GD is used to collect data.
  }\label{Infer_FHop}
\end{figure}
\end{center}
\end{widetext}

\begin{figure}
\centering
    \includegraphics[width=8.5cm]{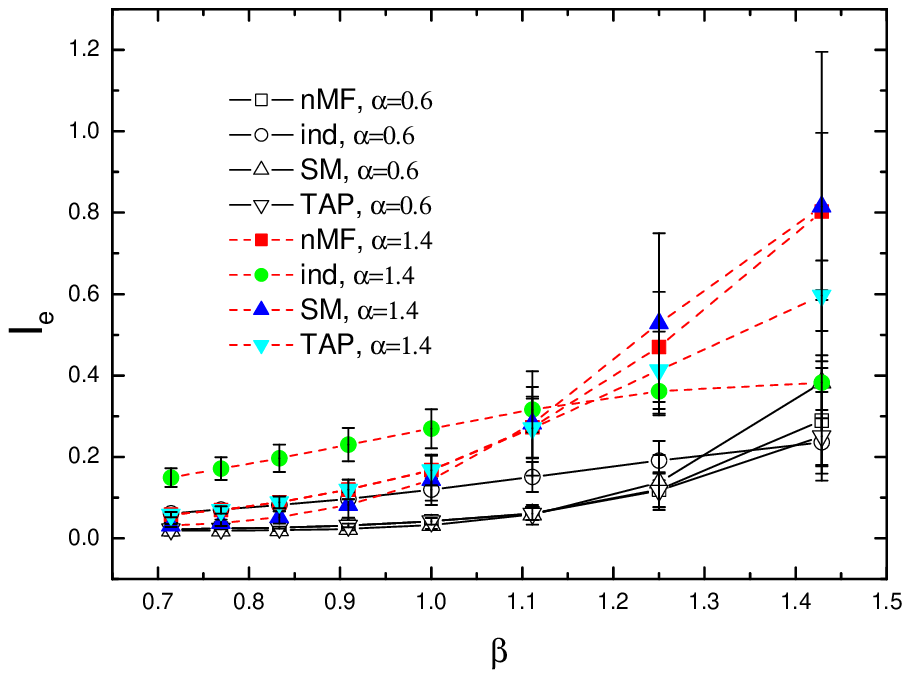}
  \caption{
    (Color online) Inference performances for the finite connectivity Hopfield network. The line linking
    each marker is a guide to the eye. The data marker is the average over $5$ samples. Two ratios $\alpha=0.6,1.4$
    are considered with the same mean degree $l=5$, system size
    $N=100$. For the low
    temperature regime, type B GD is used to collect data.
  }\label{Infer_SHop}
\end{figure}

\section{Results and Discussions}
\label{sec_result}

We restrict our analysis to the small size system, although the
phase diagram of the Hopfield model was derived in the thermodynamic
limit\cite{Amit-198501,Amit-198502,Coolen-2003,Skantzos-2004} and
the result for the small size system will be slightly different due
to the finite size effects if $N$ is not very small,  it is expected
that the recall phase, paramagnetic or spin glass phase will still
appear as $T$ and $\alpha$ vary. They can be characterized by two
order parameters, the overlap between the network configuration and
the $\mu$-th pattern
$m^{\mu}=\frac{1}{N}\sum_{i=1}^{N}\xi_{i}^{\mu}m_{i}$, and the mean
squared magnetization $q=\frac{1}{N}\sum_{i=1}^{N}m_{i}^{2}$.
Scatter plots comparing the inferred couplings with the true ones
for the fully-connected network are presented in
Fig.~\ref{SG_scatterplot}. nMF, TAP and SM give very good inferences
of couplings while ind shows a relatively high error. The part below
the line underestimates the true couplings and the part above
overestimates the true ones. As shown in Fig.~\ref{SG_scatterplot},
nMF and TAP show a similar behavior of estimating the true
couplings. They always underestimate the true negative couplings but
overestimate the positive ones. However, SM behaves conversely. As
expected, in this fully-connected network, the independent pair
approximation fails to recover the couplings within a desired
accuracy. The performances versus the system size are illustrated in
Fig.~\ref{Infer_FHop} (a). It turns out that the inference errors
increase with the network size for nMF, SM and TAP in the
paramagnetic phase. Additionally, SM performs better than nMF and
TAP. The bad predictor ind shows large fluctuations from sample to
sample. Fig.~\ref{Infer_FHop} (b) presents the reconstruction
performance versus the memory load at the low temperature. SM
behaves badly with large fluctuations, even inferior to ind for
certain memory loads. However, it should be noted that when just one
single pattern is stored, all mean field schemes perform relatively
better and give $I_{e}\simeq 0.03$. It is due to the fact that there
exists only one attraction for the retrieval dynamics, therefore it
is easy to detect the equilibrium state of the network which
contains all the information about the couplings. The reconstruction
performances against the memory load in the paramagnetic phase are
shown in Fig.~\ref{Infer_FHop} (c). As the memory load increases,
the algorithms behave worse and their inference errors show large
standard deviations. However, in the paramagnetic phase, all the
algorithms but ind give very small inference errors especially at
small memory loads. Given $\alpha$ and $N$, we report the inference
performance against temperature in Fig.~\ref{Infer_FHop} (d). In our
numerical simulations, we find that in the low temperature region,
some approximations, e.g., TAP, fail due to the high magnetizations
(very close to $1$ or $-1$) or extremely small correlations ($\sim
\mathcal{O}(10^{-4}$)) computed from GD. At the same time, the
determinant of the correlation matrix $\mathbf{C}$ is nearly equal
to zero. Therefore the algorithms are unable to extract the
couplings. These results are not shown in Fig.~\ref{Infer_FHop} (d).
For instance, we do observe the final retrieval of the stored
pattern in a simulation with $T=0.4, N=100, P=3$, nevertheless, only
nMF is successful but leads to a high inference error $\sim0.199$
and all other methods fail. In this case, given many stored
patterns, if the first pattern is assumed to be retrieved, i.e., the
overlap $\frac{1}{N}\sum_{i=1}^{N}\xi_{i}^{1}\sigma_{i}=1$, then the
energy function equation ~(\ref{Hop-H}) can be decomposed into two
terms as $\mathcal{H}=-\frac{1}{2N}\sum_{\mu\neq
1}^{P}\left(\sum_{i}\xi_{i}^{\mu}\sigma_{i}\right)^{2}-\frac{1}{2N}\left(\sum_{i}\xi_{i}^{1}\sigma_{i}\right)^{2}$,
and the second term is relevant thus
$\mathcal{H}=-\sum_{i}h_{i}\sigma_{i}$ where
$h_{i}=\frac{\xi_{i}^{1}}{2N}\sum_{j\neq i}\xi_{j}^{1}\sigma_{j}$,
which implies that in the recall phase, the information about
couplings is lost and the local field contains only the information
about the retrieved pattern. The same failure occurs also in the
spin glass phase, for example, the best performance for one single
sample is given by TAP, $I_{e}\sim0.129$ with $T=0.4, \alpha=0.13,
N=100$. As shown in Fig.~\ref{Infer_FHop} (d), the inference
performance becomes worse as the temperature decreases. With many
embedded patterns, we will lose the information about couplings even
if the retrieval successes, as explained above. On the other hand,
the free energy landscape becomes complex at low temperatures and it
is difficult to detect the equilibrium properties of the system,
i.e., the emergence of spurious minima will trap the Glauber
dynamics and it requires much longer GD steps to reduce the noise of
estimates of the magnetizations and correlations which affects the
inference results a lot. For the current system, ergodicity breaking
sets in at very low temperature where the replica symmetric
assumption was shown to be invalid~\cite{Tokita-94}, then the
magnetizations and correlations lose their physical meanings in that
they are usually defined in a single state, therefore reconstruction
algorithms capable of dealing with the case of multi-state are
necessary. Finally, it should be mentioned that in the low
temperature region, all mean field schemes show large
sample-to-sample fluctuations and their predictions are unreliable.

Inference performances of the reconstruction algorithms on the
sparse Hopfield network are also considered and shown in
Fig.~\ref{Infer_SHop}. At the low temperature, some mean field
schemes fail therefore the results are not presented. It should be
noted that ind performs well to reconstruct the finite connectivity
network, consistent with the fact that the independent pair
approximation is reasonable in the sparse network. Particularly in
the low $\alpha$ regime ind shows a very good performance, and at
the low temperature, it even outperforms other existing mean field
schemes although the inference errors are relatively large. In the
high temperature region, all mean field schemes perform better with
low $\alpha$ than high $\alpha$. At temperatures larger than $1.0$,
SM does a better job than other algorithms, as also observed in the
fully-connected case. When we decrease the temperature where the
system stays, we are confronted with the failure of all mean field
schemes as appears in the case of fully-connected network, however,
when the algorithm works well, the predicted couplings between
unconnected neurons are very small compared to those between neurons
which are actually interconnected, therefore one can reconstruct the
network using an appropriate cutoff. On the other hand, it shows
clearly that the sample-to-sample fluctuations become larger and
larger as the temperature decreases.

\section{Conclusions}
\label{sec_Con}

In this work, we have tested performances of four fast mean field
schemes for inferring couplings of Hopfield networks. We find that
nMF, SM and TAP show better performances than ind in paramagnetic
phase, and their performances degrade with the increasing network
size and memory load. The inference error achieved by SM shows large
sample-to-sample fluctuations and becomes inferior to ind in a
certain range of memory load as the system is presented in the low
temperature region. When there is one single pattern stored in the
network, all mean field schemes are able to do a good job, whereas,
as many patterns are embedded, the free energy landscape becomes
complex and even if one of embedded patterns is found, the
information about couplings of the network is inevitably lost and
the algorithms fail to reconstruct the network. Similarly, in the
spin glass phase, due to the emerging spurious minima, it is
difficult to detect the equilibrium states and the estimated
magnetizations and correlations become unreliable. The message
passing algorithms proposed in Ref.~\cite{Mezard-08} may offer hints
for improving the performance of network reconstruction. Our work
implies that as a direct problem, the recall phase is favored, while
the unfavored paramagnetic phase is in turn most useful for the
reconstruction of the network, which is particular for the Hopfield
network where couplings are constructed according to the Hebb's rule
and the stored patterns are in fact random and uncorrelated. For the
sparse network, when we decrease the temperature and this is
necessary for the network to recall one of stored patterns, all
reconstruction algorithms deteriorate and ind shows a relatively
better performance. Whether in the fully-connected case or in the
sparse case, the performances achieved by the reconstruction
algorithms show larger and larger sample-to-sample fluctuations with
decreasing temperatures and the predictions become no more accurate.

The analysis of different approximations on inferring couplings of
Hopfield network is kept to in our work, however, more analyses on
the more realistic networks, such as neural
networks~\cite{Tang-2008}, gene regulatory networks~\cite{Lezon-06}
and other biological networks~\cite{Vert-08}, are urgently required.
This subject of ongoing research will shed light not only on reasons
why some algorithms fail to reconstruct the network but also on the
further mechanisms we need to develop novel efficient network
reconstruction algorithms for practical data analysis.

\section*{Acknowledgments}

The author is grateful to Pan Zhang and Haijun Zhou for valuable
discussions. The present work was supported by the National Science
Foundation of China (Grant No. 10774150) and by the National Basic
Research Program (973-Program) of China (Grant No. 2007CB935903).



\end{document}